\documentclass[11pt, a4paper]{article}
\usepackage{amsmath}
\usepackage{amssymb}
\usepackage[dvips]{graphics}
\usepackage[dvips]{graphicx}
\usepackage{soul}

\numberwithin{equation}{section}
\usepackage{color}

\usepackage{amsmath}
\usepackage{amsfonts}
\usepackage{graphicx}
\usepackage{bm}
\usepackage{amssymb}
\usepackage{mathrsfs}
\usepackage{pstricks,pst-node}
\usepackage{amsxtra}
\usepackage{bbm}
\usepackage{paralist}
\usepackage{color}

\usepackage[symbol]{footmisc}

\newcommand{\diag}{\textup{diag}}

\DeclareMathOperator{\Tr}{Tr}

\DeclareMathAlphabet{\mathbfi}{OT1}{cmr}{bx}{it}
\DeclareMathAlphabet{\mathpzc}{OT1}{pzc}{m}{it}

\newcommand{\benumerate}{\begin{enumerate}}
\newcommand{\eenumerate}{\end{enumerate}}

\newcommand{\bitemize}{\begin{itemize}}

\newcommand{\eitemize}{\end{itemize}}

\newcommand{\der}[2]{\frac{\partial #1}{\partial #2}}

\newcommand{\dersec}[2]{\frac{\partial^{2} #1}{\partial #2^{2}}}

\newcommand{\dermixd}[3]{\frac{\partial^{2} #1}{\partial #2 ~\partial #3}}

\newcommand{\ovl}[1]{\overline{#1}}

\newcommand{\av}[1]{\langle #1 \rangle}

\begin{document}

\title{Exact equations of state for nematics}

\author{Francesco Giglio$^{\;a)}$, Giovanni De Matteis$^{\;b)}$ and Antonio Moro$^{\;a)}$}

\date{}

\maketitle
    
\begin{center}
{\small $^{a)}$ Department of Mathematics, Physics and Electrical Engineering \\ University of Northumbria at Newcastle, UK \\
$^{b)}$ Istituto di Istruzione Secondaria Superiore ``V. Lilla" \\ MIUR - Italian Ministry of Education and Research, Francavilla Fontana (BR), Italy}\footnote[1]{Contacts: francesco2.giglio@northumbria.ac.uk, dematt73@yahoo.it, antonio.moro@northumbria.ac.uk}
\end{center}



\begin{abstract}
\noindent We propose a novel approach to the solution of nematic Liquid Crystal models based on the derivation of a system of nonlinear wave equations for order parameters such that the occurrence of uniaxial and biaxial phase transitions can be interpreted as the propagation of a two-dimensional shock wave in the space of thermodynamic parameters. We obtain the exact equations of state for an integrable model of biaxial nematic liquid crystals and show that the classical transition from isotropic to uniaxial phase in absence of external fields is the result of a van der Waals type phase transition, where the jump in the order parameters is a classical shock generated from a gradient catastrophe at a non-zero isotropic field. The study of the equations of state provides the first analytical description of the rich structure of nematics phase diagrams in presence of external fields.\\

\vspace{.4cm}

\noindent Keywords: Nematic Liquid Crystals $|$ Integrability $|$ Phase Transitions $|$ Biaxiality $|$
\end{abstract}


\section{Introduction}

Liquid crystals, universally known as the material of modern displays, have been originally discovered in a completely different context back in 1888 by the botanist F. Reinitzer, who was first puzzled by the unusual optical properties of cholesterol compounds extracted from carrots and by the observation of the existence of {\it two}
melting points. It was O. Lehmann who realised that these {\it crystals that flow}  were indeed a new state of matter, giving them the name of Liquid Crystals \cite{Sluckin}. Indeed, it took about a century before the concept of Liquid Crystals would turn from a mere curiosity, although interesting and fundamental, to something of high technological impact. On a fundamental level, as observed by D. Dunmur and T. Sluckin in their history of Liquid Crystals \cite{Sluckin}, if one considers that there are no free standing plasmas on Earth, Liquid Crystals can be considered in full right as fourth terrestrial state of matter after gases, liquids and solids.

Nowadays, the physics of Liquid Crystal embraces a much wider class of substances, generally referred to as {\it soft matter} \cite{kleman_lavrentovich}, including all those systems ({\it mesophases}) whose macroscopic properties are intermediate between isotropic liquid and solid states.
Hence, the study of Liquid Crystal models is of interest for their potential technological applications as well as for being paradigmatic examples of mesoscopic systems, a test-bed for the rigorous analysis of emerging collective and complex phenomena.

In general terms, Liquid Crystals appear as a liquid substance that possesses a microstructure as a result of their molecular anisotropy. Anisotropy of molecules allows to define an orientational order and associated {\it order parameters}. Order parameters are quantities defined at the macroscopic level that acquire particular significance when collective behaviours occur in the system. Orientational order is important for example in magnetic systems where the collective behaviour of spin particles determines, below the critical temperature, the ferromagnetic phase. 

Liquid crystals are instead made of molecules with full or partial positional freedom, so that they can flow while preserving a certain orientational structure.
More specifically, so-called thermotropic {\emph{nematic}} Liquid Crystals exhibit orientational order only. In loose terms, nematics can be thought of as {\emph{rod--like}}  molecules that move randomly such that below a certain critical temperature, their molecular symmetry axis tends to align along a specific direction. The variable used to specify the orientation of a molecule is called {\emph{molecular director}}. The macroscopic state where molecular directors point on average in the same direction is called {\emph{uniaxial}} phase and the average direction defines the {\emph{director}} of the phase.  If further molecular anisotropy (lower molecular symmetry) is allowed, as in the case where the section transversal to the molecular axis does not possess rotational symmetry, an additional orientational degree of freedom can be introduced to define {\emph{biaxial}} Liquid Crystals. Biaxiality, that is the simultaneous orientation along two mutual orthogonal axes, is one of the most intriguing properties of Liquid Crystals being the subject matter of numerous modern theoretical studies~(see e.g. \cite{rexpbiall,rb01,radd00,pisapv08,padd01,bmorz}).

%

%
The existence of ordered phases combined with their liquid state is what makes Liquid Crystals highly flexible and unique for their effective deployment in the construction of displays and electronic devices. These electronic devices typically operate based on the interaction between Liquid Crystals and external electrical and magnetic fields.
%
External fields can indeed be used to control ordered phases in Liquid Crystal, determine changes of their physical properties, induce phase transitions and tune critical parameters~\cite{dunmur88, longa_external, external_nababrata, mukherjee2013external}. 

Methods for studying thermodynamic phase transitions in Liquid Crystals span from mean--field (or molecular field) statistical mechanics to Landau theory, from variational methods and classical mechanics to Monte Carlo simulations.
In this work, we propose an approach to the thermodynamics of Liquid Crystals based on the theory of nonlinear integrable conservation laws, intended as a set of nonlinear PDEs of hydrodynamic type. We show that order parameters satisfy suitable nonlinear wave equations and phase transitions are interpreted as the occurrence of shocks in the wave dynamics. This method for solving thermodynamics relations and construct equations of state has been introduced in \cite{GenoveseBarra, DeNittis, BarraMoro} and extended  to more general class of fluid and magnetic models in \cite{Guerra,Agliari,Landolfi,Fachechi}. 

We consider a discrete version of the Maier-Saupe model~\cite{maier-saupe1, maier-saupe2, dispersion_london1, dispersion_london2, dispersion_london3, rPV01, rPV02, rPV04}, referred to as MS6 model \cite{Nascimento} with the addition of external fields. In the MS6 model molecules are represented by elongated cuboids (biaxial molecular symmetry) whose only admissible configurations are such that molecular directors are restricted to Cartesian directions. 
We use methods of integrable conservation laws to derive explicit equations of state for this model. We show that, in presence of external fields, the system admits phase transitions of van der Waals type and provide an interpretation of the jump singularity of order parameters observed when  external fields vanish in terms of the shock dynamics of a nonlinear wave.
This model, although relatively simple, reproduces all fundamental features of phase transitions in nematics. The proposed approach is rigorous and leads to the exact equations of state for a biaxial liquid crystal model with external field. 

The paper is organised as follows: in Section 1, we introduce the discrete version of the Maier-Saupe model with external fields and the finite-size differential identities for the partition function; Section 2 is dedicated to the derivation of the equations for the free energy and order parameters in the thermodynamic limit for general biaxial molecules; Section 3 provides a detailed study of the equations of state and their singularities. Phase diagrams are discussed in Section 4. A summary of results obtained and further outlook is the subject of the final Section 5.

\section{Discrete Maier-Saupe model with external fields}
Let us consider a system of $N$ interacting Liquid Crystals molecules shaped as elongated cuboids whose molecular directors $\vec{n}_1,\vec{n}_2$ and $\vec{n}_3$ are unit vectors parallel to their principal axes. Introducing the tensors (see e.g. \cite{rPV01}) 
\begin{equation}
{\bf q} = \vec{n}_1 \otimes \vec{n}_1 - \frac{1}{3} { \bf I} \qquad {\bf b} = \vec{n}_{2} \otimes \vec{n}_{2} - \vec{n}_{3} \otimes \vec{n}_{3}  
\end{equation}
where ${\bf I}$ is the $3 \times 3$ identity matrix and the linear combination 
\[
\ovl{{\bf q}} = {\bf q} + (\Delta/3) {\bf b}
\]
we consider the Hamiltonian of the form
\begin{equation}
\label{eq:H0}
H_{0} = - \frac{9 \mu}{2N} \sum_{i,j} \ovl{{\bf q}}_{i}  \cdot \ovl{{\bf q}}_{j}
\end{equation}
where $\ovl{{\bf q}}_i$ is associated to the state of the $i-$th molecule and by definition  $ {\bf a} \cdot {\bf b} := \Tr{({\bf a b})}$, where $\Tr{}$ is the trace operator. Summation indices $i$ and $j$ run from $1$ to $N$ and $\mu$ is the mean field coupling constant. For convenience, we have included self-interaction terms corresponding to $i=j$. This choice will not affect the result as it corresponds to a shift of the energy reference frame by a constant. 
Using the assumption that allowed configurations are such that molecular directors are parallel to Cartesian axes, the Hamiltonian (\ref{eq:H0}) can be written as follows
\begin{equation}
H_0=- \frac{\mu}{2 N} \sum_{i,j} {\bf \Omega}_{i} \cdot {\bf \Omega}_{j} = -\frac{\mu}{2 N} \sum_{i,j} \sum_{k,l \in \{1,2\}} c_{kl} \Lambda_{i}^{k} \Lambda_{j}^{l}  \, ,
\end{equation}
where $c_{kl}=1 + \delta_{kl}$, with $\delta_{kl}$ the Kronecker delta and ${\bf \Omega}_{i}$ are traceless diagonal matrices of the form
\begin{equation}
\label{MS6defintionOmega}
{\bf \Omega}_{i} = \diag(\Lambda^{1}_i,\Lambda^{2}_i, -\Lambda^{1}_i-\Lambda^{2}_i)
\end{equation}
given by the six possible orientational states of each molecule ${\bf \Omega}^{(k)}$, with $k=1,\cdots,6$. In particular, we have ${\bf \Omega}^{(1)} = \diag(-1+\Delta, -1-\Delta,2)$, and ${\bf \Omega}^{(2)}, \dots,{\bf \Omega}^{(6)}$ are obtained by considering all permutations of diagonal entries of ${\bf \Omega}^{(1)}$.
Introducing the quantities $M^k= \sum_{i} \Lambda_{i}^{k}/N$ with $k=1,2$, the Hamiltonian $H_{0}$ reads as follows
\begin{equation}
\label{ms6hamiltonian}
H_0= -\mu N \left ((M^{1}) ^{2} + M^1 M^2 + (M^{2})^{2}\right).
\end{equation}
Let us assume that individual molecules interact with the external field ${\boldsymbol \epsilon}= \diag \left (\epsilon_1, \epsilon_2,\epsilon_3 \right)$ where the interaction potential is of the form $H_{ex} = - 3 \sum_{i} \boldsymbol{\epsilon} \cdot \ovl{{\bf q}}_{i} = - \sum_{i} \boldsymbol{\epsilon} \cdot {\bf \Omega}_{i}$.
Hence, using the notation introduced above, we can write
\begin{equation}
\label{eq:Hex}
H_{ex} = - N \left( \epsilon_{13} M^1 + \epsilon_{23} M^2 \right)
\end{equation}
where $\epsilon_{k3} = \epsilon_k - \epsilon_3 $, with $k = 1,2$. If $\epsilon_{i} = \epsilon_{j}$ for some $i\neq j$ the external field is said to be {\it isotropic}.  Hence, the full Hamiltonian for the MS6 model with external fields is $H = H_{0} + H_{ex}$. Its thermodynamic properties are standardly described by the partition function associated with the Gibbs distribution
\[
Z_{N} = \sum_{\{ \ovl{{\bf q}} \}} \exp (-\beta H)
\]
where the summation refers to all possible configurations of states $\ovl{{\bf q}}_i$ and $\beta = 1/T$ with $T$ the absolute temperature. Introducing the rescaled coupling constants $t:= \beta \mu$, $x :=  \beta \epsilon_{13}$ and $y :=  \beta \epsilon_{23}$ the partition function reads as
\begin{equation}
\label{eq:partition}
Z_{N} = \sum_{\{ \ovl{{\bf q}} \}} e^{N \left \{ t \left [(M^1)^2 + M^1 M^2 + (M^2)^2  \right ] + x M^1 + y M^2 \right \}} ~.
\end{equation}
We note that with this notation the external field is isotropic if  either $x = y$ or $x=0$ or $y=0$.
In the following, similarly to the case of spin systems~\cite{Guerra} and van der Waals type models \cite{BarraMoro}, we look for a differential identity satisfied by the partition function~(\ref{eq:partition}) and calculate the associated initial condition. 
We observe that the partition function~(\ref{eq:partition}) identically satisfies the $2+1$-dimensional PDE
\begin{equation}
\label{eq:heat}
\der{Z_{N}}{t} = \frac{1}{N} \left ( \dersec{Z_{N}}{x} + \dermixd{Z_{N}}{x}{y} + \dersec{Z_{N}}{y} \right).
\end{equation}
Note that this equation can be transformed into the heat equation of heat conductivity $1/N$ by a linear change of variables.
The associated initial condition, $Z_{N}(x,y,t=0) = Z_{0,N}(x,y)$, is provided by the value of the partition function of the model for non mutual interacting molecules.
Given that the exponential is linear in the variables $M^1$ and $M^2$, the initial condition can be evaluated by recursion and gives the following formula
\begin{equation}
\label{eq:partition0}
Z_{0,N} = \left( \sum_{l=1}^{6} e^{x\Lambda^{1,l} + y \Lambda^{2,l}} \right)^N
\end{equation} 
where the index $l$ labels the pairs $(-1+\Delta,-1-\Delta)$, $(-1+\Delta,2)$, $(-1-\Delta,2)$, $(-1-\Delta,-1+\Delta)$, $(2,-1+\Delta)$ and $(2,-1-\Delta)$. The exact solution to the equation~(\ref{eq:heat}) for a given number of molecules $N$ can be  formally obtained by separation of variables by using as a basis the set of exponential functions obtained by expanding the $N-$th power at the r.h.s. of equation (\ref{eq:partition0}). The solution reads as
\begin{equation}
\label{finiteNZ}
Z_N = \sum_{\vec{k}} B_{\vec{k}} \; A_{\vec{k}}(t)  \exp \left (x \lambda^{1}_{\vec{k}} + y \lambda^{2}_{\vec{k}} \right)
\end{equation}
where $\vec{k} = (k_{1},\dots,k_6)$ is a multi-index such that $k_l = 0,\dots,N_l$ with $N_1=N$, $N_l = N_{l-1}-k_l$ for $l=2,\dots,5$, $k_6 = N_6$, $\lambda^{s}_{\vec{k}}  =\sum_{l=1}^6 \Lambda^{s,l} k_{l}$, $s=1,2$ and
\[
B_{\vec{k}} =\prod_{l=1}^6 \binom{N_l}{k_l}, \quad A_{\vec{k}} = \exp \left \{ \frac{t}{N} \left [  \left(\lambda^{1}_{\vec{k}} \right)^2  +\lambda^{1}_{\vec{k}} \lambda^{2}_{\vec{k}} +\left(\lambda^{2}_{\vec{k}} \right)^2    \right]   \right \}.
\]
Let us define the order parameters  $m_{N}^1$ and $m_{N}^{2}$, as the expectation values of, respectively, the {\it order parameters} $M^1$ and $M^2$, i.e.
\begin{equation}
m^k_N := \av{M^k} = \frac{1}{Z_N} \sum_{\{ \ovl{{\bf q}} \}} M^k e^{-\beta H}, \qquad k=1,2.
\end{equation}
Introducing the free energy per particle $
{\cal F}_{N}:= (1/N)\log Z_N$,
the order parameters can be calculated by direct differentiation as follows
\begin{equation}
\label{eq:orderpar}
m^1_N = \der{{\cal F}_{N}}{x}, \qquad  m^2_N = \der{{\cal F}_{N}}{y}.
\end{equation}
Above formulae, although explicit, are effective if $N$ is sufficiently small due to slow binomial convergence.  In section below, we derive the equations of state in thermodynamic (large $N$) regime  via direct asymptotic evaluation of the equation~(\ref{eq:heat}).

\section{Thermodynamic limit and equations of state}
The thermodynamic limit is defined as the regime where the number of particles $N$ is large, i.e. $N \to \infty$. Assuming that the free energy admits the expansion of the form ${\cal F}_{N} = F + O \left (1/N \right)$ and using the equation (\ref{eq:heat}) we obtain, at the leading order, the following Hamilton-Jacobi type equation
\begin{equation}
\label{eq:HJF}
\der{F}{t} =  \left( \der{F}{x} \right)^{2} +  \der{F}{x} \der{F}{y} + \left(\der{F}{y} \right)^{2} .
\end{equation}
A similar asymptotic expansion for the order parameters $m^k_N = m^k + O(1/N)$ implies the relations 
\[
m^1 = \der{F}{x}, \qquad m^2 = \der{F}{y}.
\]
Equation (\ref{eq:HJF}) is completely integrable and can be solved via the characteristics method. In particular, the solution can be expressed via the free energy functional
\begin{equation}
\label{eq:Fsol}
F = x m^1 + ym^2 +t \left[(m^1)^2 + m^1 m^2 +(m^2)^2\right]  - W(m^1,m^2)
\end{equation}
where $m^1$ and $m^2$ are stationary points of the free energy
\[
\der{F}{m^1} = 0 \qquad \der{F}{m^2} = 0
\]
and therefore solutions of the following system of equations
\begin{equation}
\label{eq:ESgen}
x + t (2 m^1 + m^2) = \der{W}{m^1}, \quad y + t (m^1 +2 m^2) = \der{W}{m^2}.
\end{equation}
The function $W(m^1,m^2)$, as discussed below, can be uniquely fixed via the initial condition $F_0 =F(x,y,t=0)$.
System (\ref{eq:ESgen}) gives the set of equations of state for the MS6 model. Hence, phase transitions underlying the model can be studied through the analysis of critical points for the set of state equations (\ref{eq:ESgen}).

Similarly to the thermodynamic models studied in~\cite{DeNittis,AntonioAnnals,Guerra,BarraMoro,Agliari}, $m^1$ and $m^2$ can be viewed as solutions to a nonlinear integrable system of hydrodynamic type where coupling constants $x$, $y$ and $t$ play the role of space and time variables respectively. In this framework, state curves associated to phase transitions correspond precisely to shock waves of the hydrodynamic flow.
In order to completely specify the equations of state (\ref{eq:ESgen}) we have to determine the function $W(m^1,m^2)$. We proceed by evaluating the equations (\ref{eq:ESgen}) at $t=0$, that is
\begin{equation}
\label{eq:ESgen0}
x(m^{1}_{0},m^{2}_{0})  = \left . \der{W}{m^1} \right |_{m^k =m^k_0}, \quad y(m^{1}_{0},m^{2}_{0}) = \left . \der{W}{m^2} \right |_{m^k =m^k_0},
\end{equation}
where $m^k_0 = m^k(x,y,t=0)$, $k=1,2$. Equations (\ref{eq:ESgen0}) show that the function $W(m^1,m^2)$ can be obtained, locally, by expressing $x$ and $y$ as functions of the order parameters $m^1$ and $m^2$ evaluated at $t=0$ and then integrating the equations (\ref{eq:ESgen0}). Indeed, observing that the initial condition for $F$ is $F_0 = {\cal F}_{N,0} = (1/N) \log Z_0$, where $Z_0$ is given in (\ref{eq:partition0}), the required functions are obtained by inverting the system
\begin{equation}
\label{minvert}
m^1_0 = \der{F_0}{x}(x,y) \qquad m^2_0 = \der{F_0}{y}(x,y).
\end{equation}
More explicitly, equations (\ref{minvert}) read as follows
\begin{equation}
\label{roots}
\sum_{l=1}^{6} \left(m^{k}_0 - \Lambda^{k,l} \right) X^{\Lambda^{1,l}} Y^{\Lambda^{2,l}} = 0 \qquad k=1,2
\end{equation}
where we have introduced the notation $X = \exp (x)$, $Y = \exp( y)$. Hence, the equations of state~(\ref{eq:ESgen}) for the MS6 model with external fields are completely determined in terms of the roots of the system of equations~(\ref{roots}). We should also emphasise that for integer values of the parameter $\Delta$ the system (\ref{roots}) is algebraic with respect to the variables $X$ and $Y$.

\section{Uniaxial equations of state with external fields} 
\label{uniaxial equations of state}
\begin{figure}[h]
\begin{center}
\includegraphics[height=5cm]{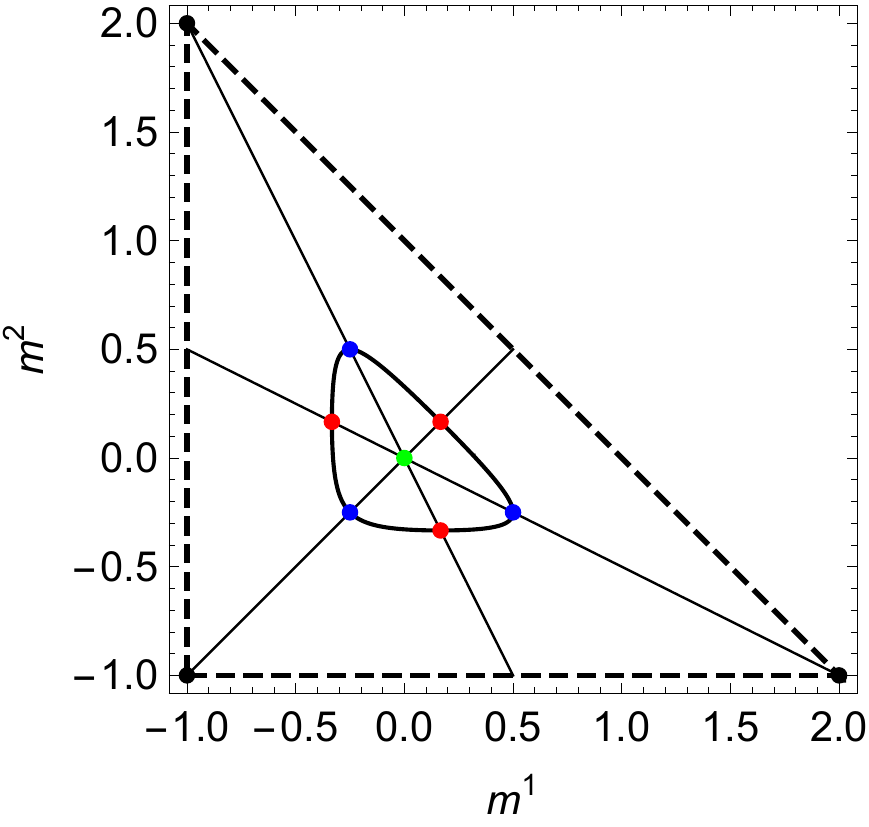}
\includegraphics[height=5cm]{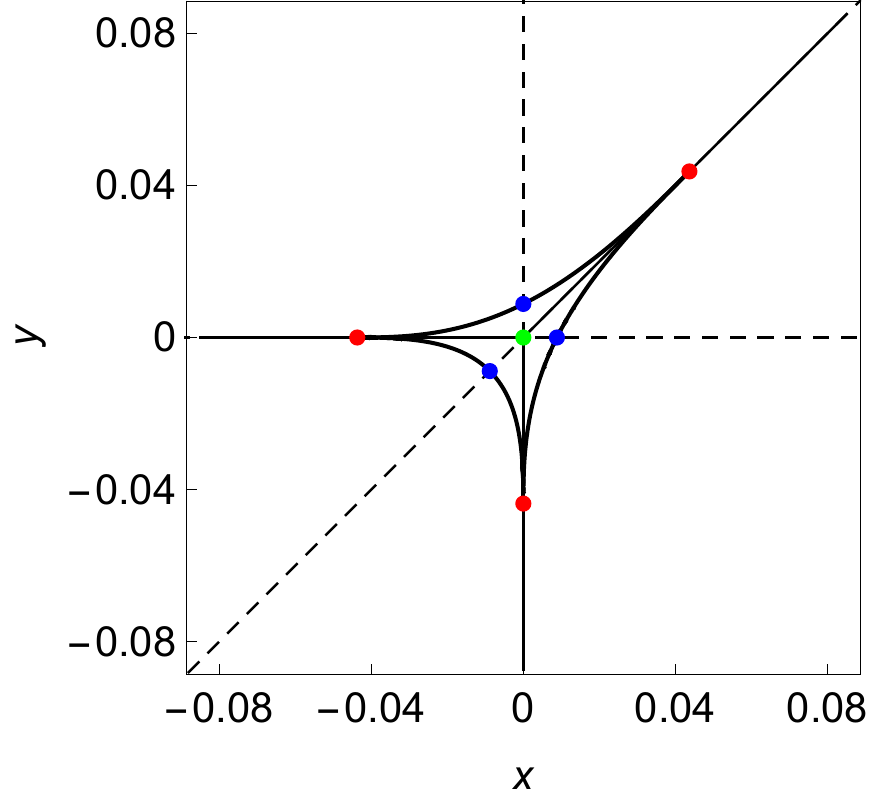}
\caption{
\label{Fig:cusps}
{\it Left:}  Critical set (locus of cusp points) constituted by  uniaxiality lines (solid thin lines) and the loop (thick line). The triangular region delimited by the dashed lines corresponds to admissible values of $m^1 $ and $m^2$. Three cusp points appear for $t > 2/9 $ from the edges of the domain and propagates each along uniaxiality lines towards the origin. Intersection with the loop (red dots) occurs at $t = 2/7 $ where six additional cusp points emerge  and propagate along the loop. 
At $t = 8/27 $ the cusp points propagating along the loop collide and annihilate in pairs at intersection with the uniaxiality lines (blue dots). For $t > 8/27 $ only three cusps along the uniaxiality lines survive, merging at the origin for $t = 1/3 $. {\it Right:} Image of the critical set via the mapping~(\ref{Omega}). The triangular region is mapped to the whole $(x,y) $ plane, uniaxiality lines corresponds to $x$ and $y$ negative semi-axes and the first quadrant bisector (thin solid lines). The loop is mapped onto the arrowhead shaped contour (thick solid line).} 
\end{center}
\end{figure}

In the present context, a uniaxial Liquid Crystal corresponds by definition to the choice $\Delta = 0$. In this case, it might be useful to think of Liquid Crystal molecules as elongated cuboids of squared section. Then, equations (\ref{roots}) give
\begin{equation}
\label{xydelta0}
x = \frac{1}{3} \log \left (\frac{m^1_0 + 1}{1-m^1_0-m^2_0} \right), \quad y = \frac{1}{3} \log \left (\frac{m^2_0 + 1}{1-m^1_0-m^2_0} \right),
\end{equation}
where $
(m^1_{0},m^2_{0})$ is an element of the set 
$$
\mathcal{T}= \left\lbrace (m^1,m^2)\in \mathbb{R}^2 | -1 < m^1< 2, -1< m^2<1-m^1 \right\rbrace 
$$
shown in Figure~\ref{Fig:cusps}. \\

Integrating the equations (\ref{eq:ESgen0}) where the functions $x(m^{1}_{0},m^{2}_{0})$ and $y(m^{1}_{0},m^{2}_{0})$ are given by (\ref{xydelta0}), we obtain the expression for the function $W$, i.e.
\begin{equation}
W = \frac{1}{3} \sum_{i=1}^{3} (1+m^i) \log (1+m^i) 
\end{equation}
with $m^3 = -m^1-m^2$.

For this choice of the function $W$, equations (\ref{eq:Fsol}) and (\ref{eq:ESgen}) provide, respectively, the free energy and equations of state for the uniaxial nematic Liquid Crystals model. 

We shall also note that the function $W$ can be expressed in terms of the invariants $I_k=(m^1)^k+(m^2)^k+(m^3)^k$  via the following formal series 
\[
W= \frac{1}{3} \sum_{k=2}^{\infty} \frac{(-1)^k}{k(k-1)}I_k \,,
\]
giving the following expression for the free energy (\ref{eq:Fsol})
\begin{equation}
F=x m^1 + y m^2 +\frac{1}{2} \left(t-\frac{1}{3} \right) I_2-\frac{1}{3} \sum_{k=3}^{\infty} \frac{(-1)^k}{k(k-1)}I_k \,.
\end{equation}
We recall that for any symmetric and traceless $3\times 3$ matrix, all $I_k$'s can be expressed as polynomials in $I_2=2\left((m^1)^2+m^1 m^2 +(m^2)^2\right)$ and $I_3=-3 m^1 m^2 (m^1 +m^2)$, known in the literature as  {\it fundamental invariants}.

Equations of state~(\ref{eq:ESgen})  allow to calculate the critical points and then the phase diagram of the uniaxial Liquid Crystals subject to external fields. The critical point can be obtained via the analysis of singularities of the family of maps $\Psi: (m^1,m^2)\in \mathcal{T} \subset \mathbb{R}^2 \to (\Psi^{1},\Psi^{2})\in \mathbb{R}^2 $ defined by the equations of state (\ref{eq:ESgen}), i.e.
\begin{gather}
\label{Omega}
\begin{aligned}
&\Psi^1:= x + (2  m^1 + m^2)t - \frac{1}{3}\log \frac{1+m^1}{1+m^3} = 0 \\ 
&\Psi^2:= y + ( m^1 + 2 m^2)t -\frac{1}{3}\log \frac{1+m^2}{1+m^3} = 0
\end{aligned}
\end{gather}
where $(m^1,m^2) \in \mathcal{T}$ and $m^3 = -m^1 - m^2$.
The variable $t$ parametrises the family of mappings.

\subsection{Singularities} The critical points of the nematic phase transition corresponds to the cusp points of the map $\Psi$. By definition, the singular sector - {\it general fold} -  of the uniaxial Liquid Crystal is given by the condition
\begin{equation}
\label{genfold}
J(m^1,m^2;t) := \der{\Psi^1}{m^1} \der{\Psi^2}{m^2} -\der{\Psi^1}{m^2} \der{\Psi^2}{m^1} = 0,
\end{equation}
where $J(m^1,m^2;t)$ is the Jacobian of the map $\Psi$. Cusp points are points of the general fold such that
\begin{align}
\label{cuspcond}
&\frac{\partial \Psi^i}{\partial m^1} w_1 + \frac{\partial \Psi^i}{\partial m^2} w_2=0, \qquad i =1,2 
\end{align}
where the vector field $w = (w_1,w_{2}) = (-\partial J/\partial m^2,  \partial J/\partial m^1)$ is tangential to the general fold. We observe that the two equations~(\ref{cuspcond}), when restricted to the general fold, are proportional to each other due to the condition~(\ref{genfold}). Hence, without loss of information one can consider only one of the two equations~(\ref{cuspcond}) along with the equation~(\ref{genfold}). In particular, equations~(\ref{genfold}) and~(\ref{cuspcond}) imply that the locus of cusp points is given by the {\it uniaxiality} lines
\begin{align}
\label{unixline}
\textup{(I)~} m^1 = m^2  \qquad \textup{(II)~} 2 m^1 + m^2 =0 \qquad \textup{(III)~}m^1 + 2 m^2 =0
\end{align}
and the closed curve of equation
\begin{equation}
\label{loop}
\textup{(IV)~}25  I_{3} = 2 I_{2}^{2} + \frac{27}{2} I_{2} - 3 \, .
\end{equation}
The critical set given by the union of uniaxiality lines~(\ref{unixline}) and the loop~(\ref{loop}) are shown in Figure~\ref{Fig:cusps} ({\it Left}).
Uniaxiality lines are the locus where the expected order tensor possesses two equal eigenvalues. Indeed, the condition $m^{i} = m^{j}$ with $i \neq j$ combined with the zero-trace condition $m^{1} + m^{2} + m^{3} = 0$ is equivalent to the set of equations~(\ref{unixline}).
Let us consider, for example, the uniaxiality line (II). The cusp exists for $t > 2/9$. Its position is
\begin{equation}
m^1(t)=\frac{1}{3 t}-1 \qquad m^2(t)=2\left(1-\frac{1}{3 t}\right),
\end{equation}
and the corresponding values of the external fields as $t$ varies, are
\begin{equation}
 x(t)=0 \qquad y(t)=1-3t + \frac{1}{3} \log (9t-2)
\end{equation}
where $\frac{2}{9} < t < \frac{1}{3}$. We should also observe that for $t=1/3$ the cusp singularity becomes unstable as the rank of the Jacobian matrix of the map $\Psi$ vanishes. Moreover, given that $y(t) < 0$ within the above range of $t$,  this means that the uniaxiality line (II) is mapped to the negative $y-$semiaxis.

Similarly, as shown in Figure~\ref{Fig:cusps}, the uniaxiality line (III) is mapped onto the negative $x-$semiaxis and the uniaxiality line (I) is mapped to the first quadrant bisecant.

\section{Phase diagrams}
\label{Phase diagrams}
The following analysis of the equations of state~(\ref{Omega}) shows that for uniaxial molecules, i.e. $\Delta = 0$, the macroscopic state of the system is uniaxial or isotropic if it is subject to an isotropic ($x=y$) or vanishing external field ($x=y=0$) and macroscopically biaxial otherwise. Consistently with the classical result, in absence of external fields the order parameter develops a jump. We provide a wave dynamics interpretation of this jump as a  confluence on the $(x,y)$ plane of three shocks that have formed at three symmetrically located points on the uniaxiality lines and travel towards the origin.

\begin{figure}[h]
\begin{center}
\includegraphics[width=4.5cm]{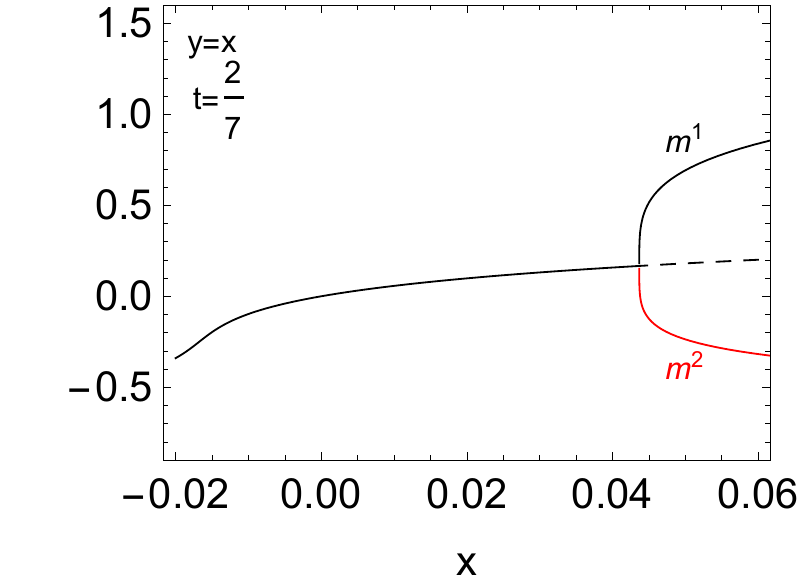}
\includegraphics[width=4.5cm]{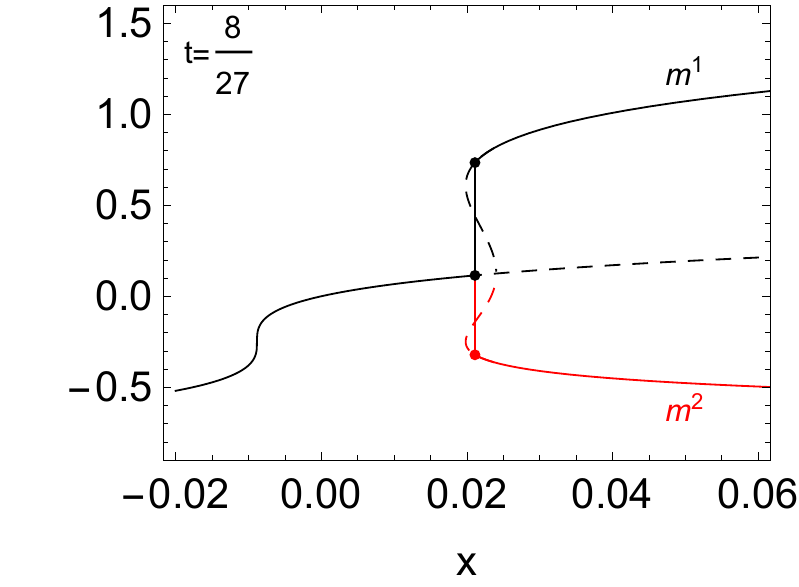} \\
\includegraphics[width=4.5cm]{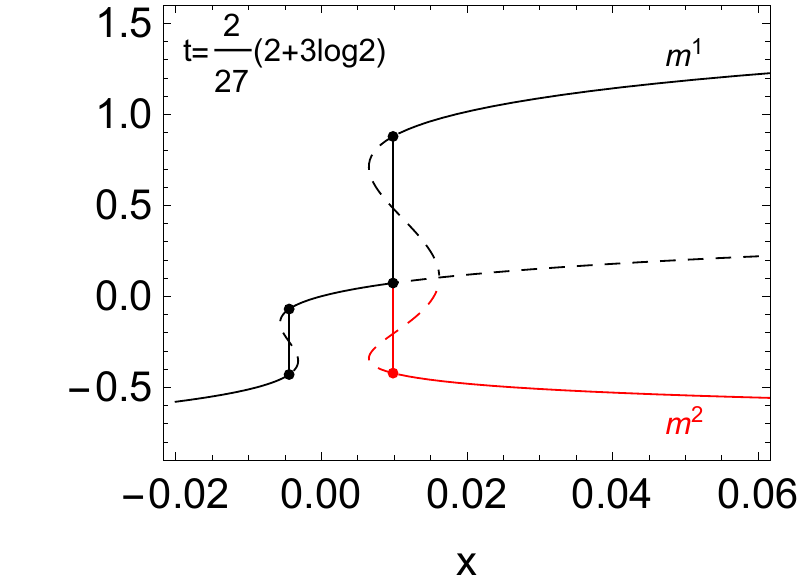}
\includegraphics[width=4.5cm]{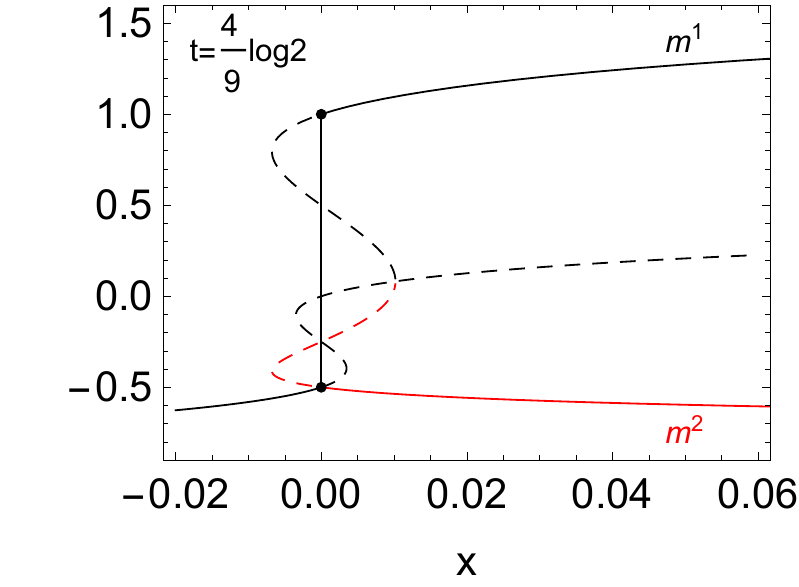}
\caption{
\label{Fig:mvsx=y}
Order parameters $m^{1}$ and $m^{2}$ along the line $y =x$ for different values of $t \geq 2/7.$} 
\end{center}
\end{figure}

\begin{figure}[h]
\begin{center}
\includegraphics[height=5cm]{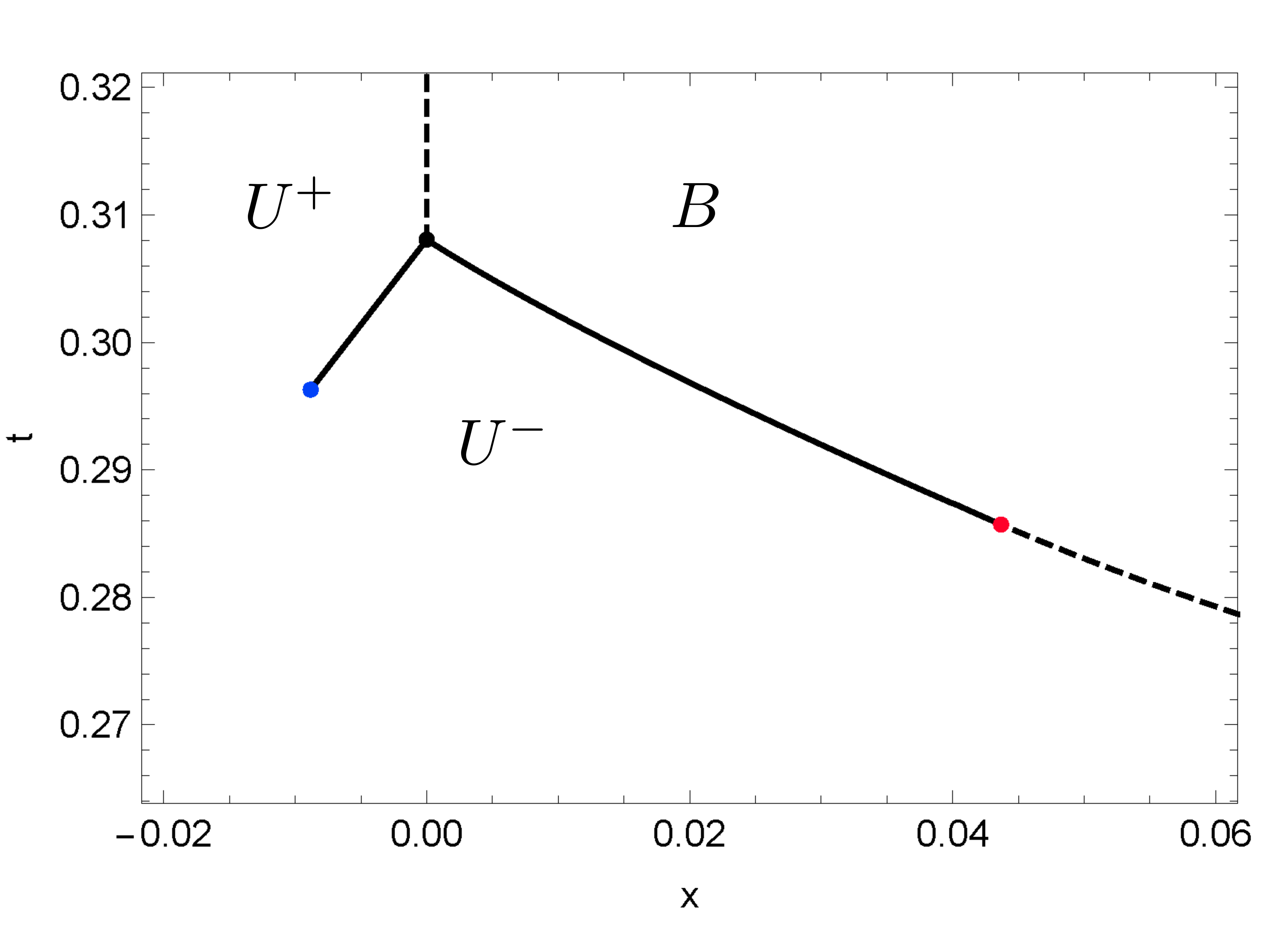}
\caption{
\label{Fig:Uniaxial}
Shock trajectories (solid lines) on the $(x,t)$ plane for $y=x$ represent the coexistence lines separating the indicated macroscopic phases, uniaxial nematic $U^{+}$, uniaxial paranematic $U_{-}$ and biaxial $B$. Dashed lines are associated to discontinuous derivatives of order parameters.
}
\end{center}
\end{figure}

\begin{figure}[h]
\begin{center}
\includegraphics[height=3.5cm]{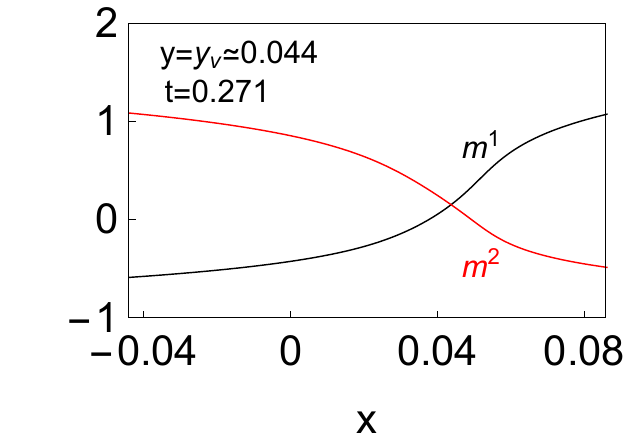} 
\includegraphics[height=3.5cm]{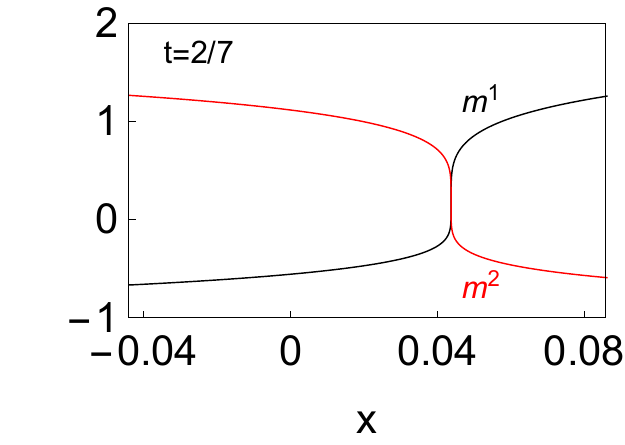} 
\includegraphics[height=3.5cm]{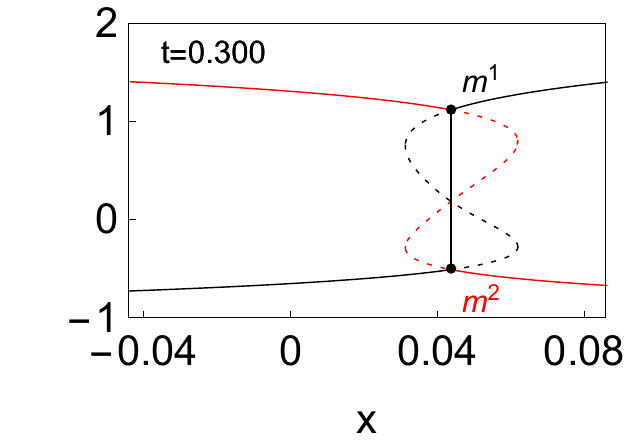} 
\caption{
\label{Fig:uniaxialjump}
Evolution of the order parameters along the line $y \simeq 0.044$ for $t <2/7$, $t  = 2/7 $  and $t > 2/7$.  The value $t = 2/7$ is the {\it critical time} where order parameters develop a gradient catastrophe. } 
\end{center}
\end{figure}

\begin{figure}[ht]
\begin{center}
\includegraphics[width=4.5cm]{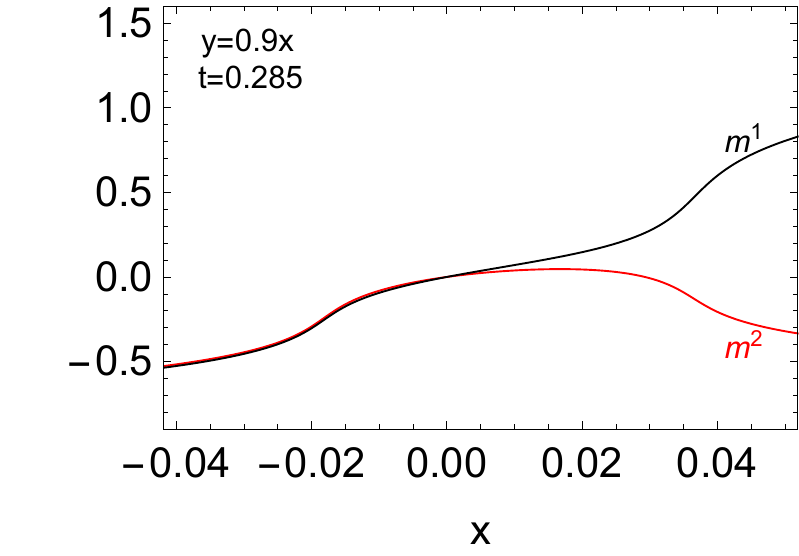}
\includegraphics[width=4.5cm]{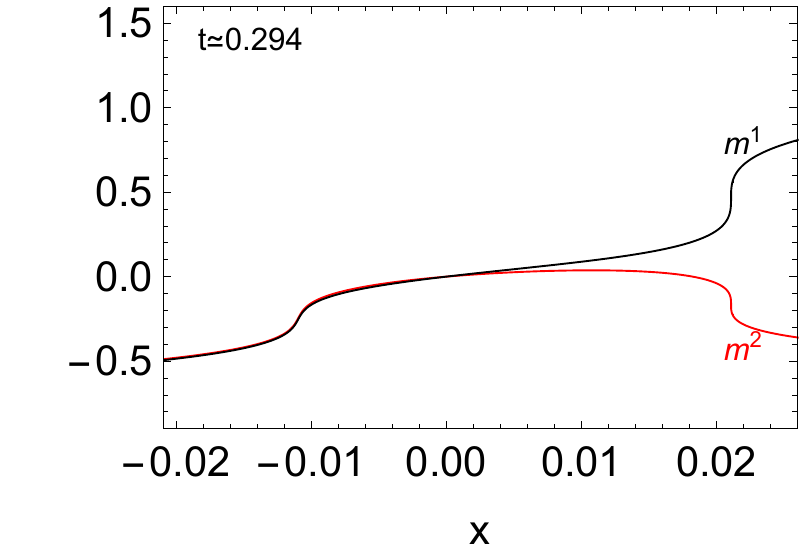} \\
\includegraphics[width=4.5cm]{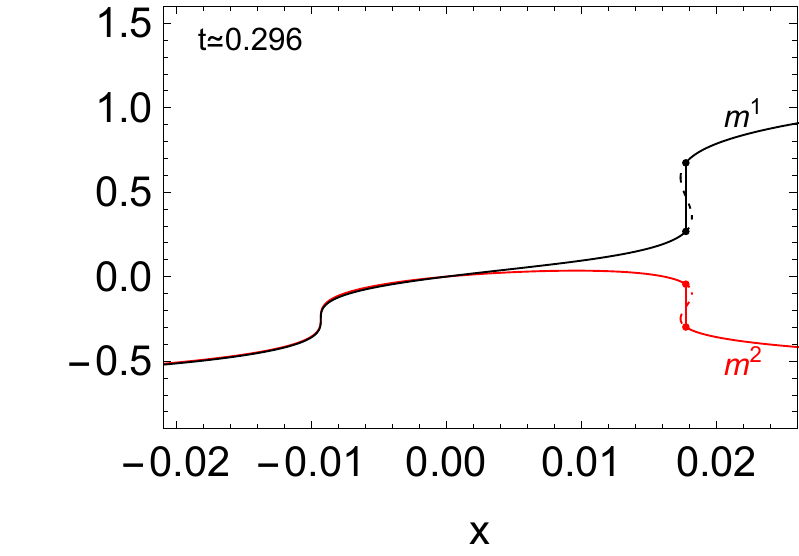}
\includegraphics[width=4.5cm]{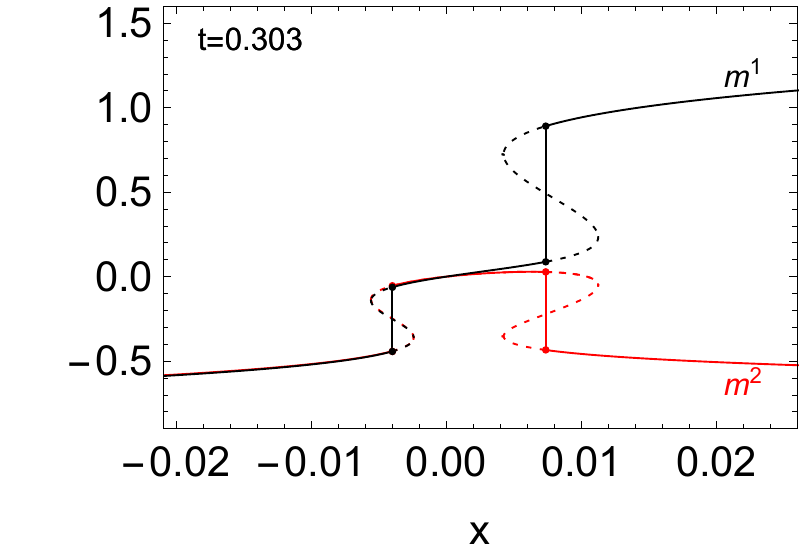}
\caption{
\label{Fig:genericline}
Order parameters along the line $y = 0.9 x$. The system is weakly biaxial for $x< 0$. Biaxiality is more pronounced for $x > 0$. At $t = 0.294$ and $t=0.296$ order parameters develop a gradient catastrophe for, respectively $x>0$ and $x<0$.} 
\end{center}
\end{figure}

\begin{figure}[h]
\begin{center}
\includegraphics[height=5cm]{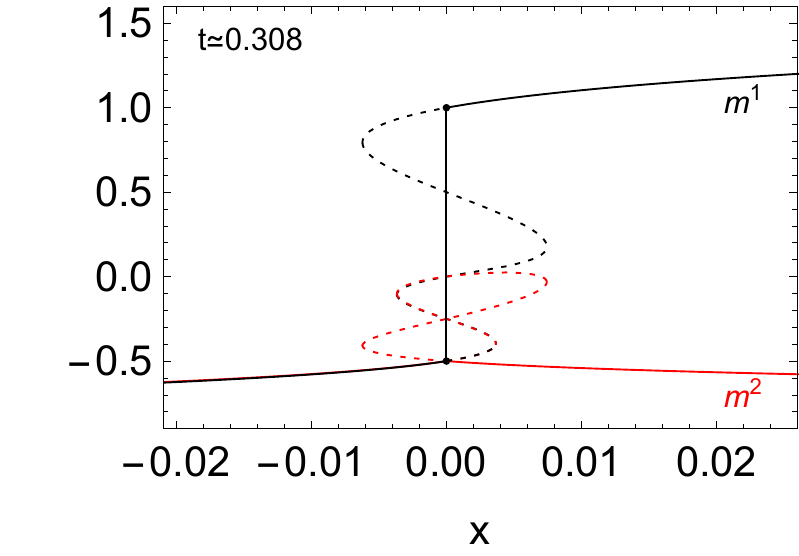}
\includegraphics[height=5.24cm]{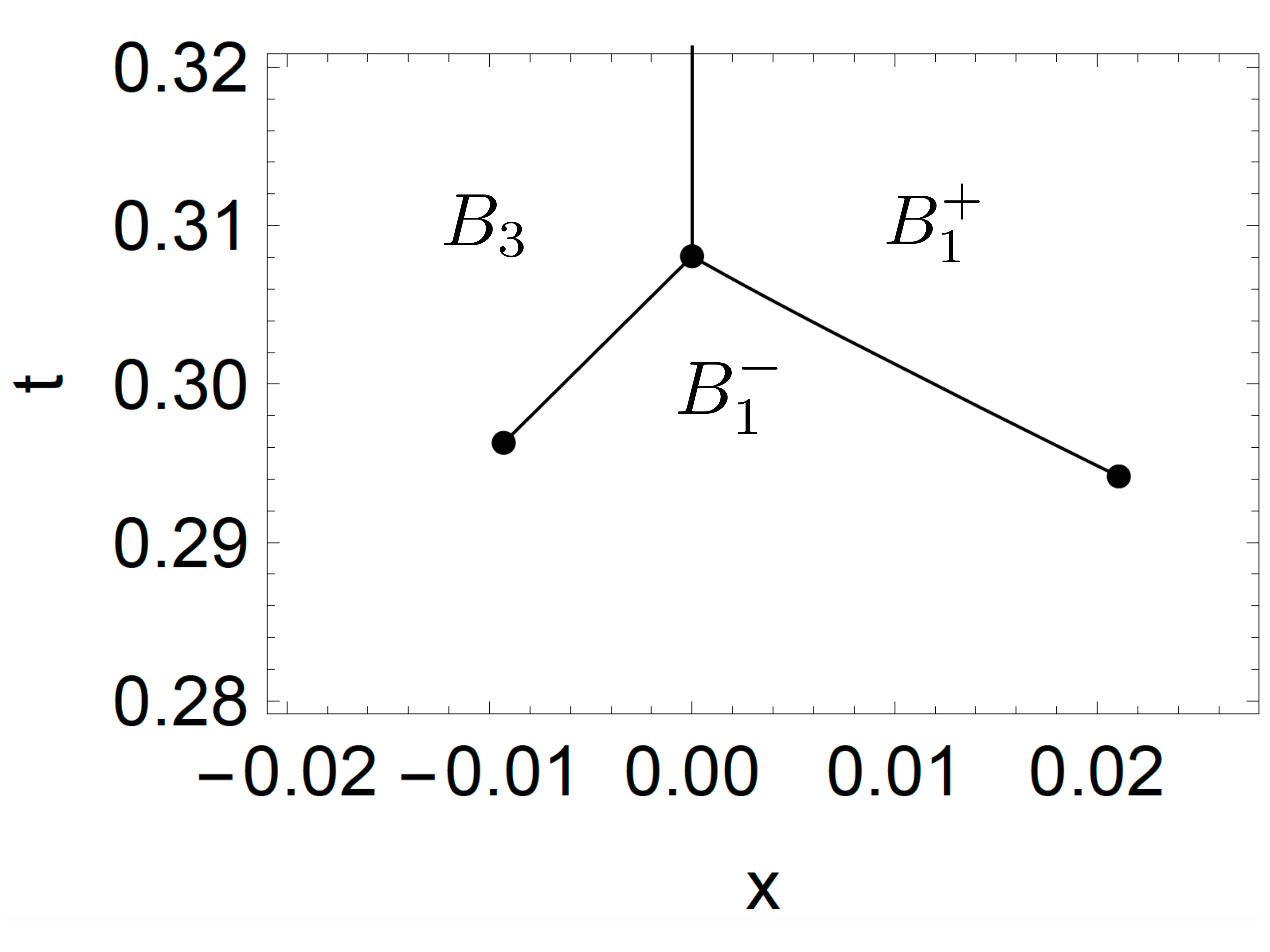}
\caption{
\label{Fig:genericline2}
({\it Left}) Confluence of shocks at the origin for $m^{1}$, $m^{2}$  and   ({\it Right}) associated ($x,t$) phase diagram along the line $y = 0.9 x$. The biaxial phase $B_3$ corresponds to molecules mostly oriented along the third Cartesian axis. A sharp transition between a lower and higher degree of order occurs along the line separating the phase $B_1^-$ and $B_1^+$ where molecules are mainly oriented along the first Cartesian axis.
} 
\end{center}
\end{figure}

The study of equations of state~(\ref{Omega}) provides a detailed description of mechanisms responsible for the occurrence of jump singularities of order parameters $m^{1}$ and $m^{2}$ and then phase transitions. Let us first consider the case of isotropic external field, for instance, such that $x=y$, i.e. $\epsilon_{1} = \epsilon_{2}$. Similar considerations hold in the cases $x=0$, i.e. $\epsilon_{1} = \epsilon_{3}$, and $y=0$, i.e. $\epsilon_{2} = \epsilon_{3}$ .  
Figures~\ref{Fig:mvsx=y} show that along the uniaxiality line $m^{1}=m^{2}$, that corresponds to $x=y$, cusp singularities, associated to a gradient catastrophe of order parameters,  occur along the first and third quadrant bisecant of the $(x,y)$ plane. If $x<0$, the cusp is an isolated point obtained from the intersection between the line $y=x$ and the loop corresponding to
 $x \simeq -0.008$.
The order parameter develops a gradient catastrophe at the isolated cusp at $t = 8/27$ and multi-valuedness at subsequent $t$'s. Hence, for $t>8/27$ the free energy admits multiple critical points meaning that the physical solution develops a jump such that the free energy maxima attain the same value. The shock travels at velocity ${\bf v}_{I}$ along $y=x$ and reaches the origin $(x,y) = (0,0)$ at  $t = (4/9) \log 2$ where it collides inelastically with the other two shocks travelling along the lines $x=0$ and $y=0$, respectively, with velocities ${\bf v}_{II}$ and ${\bf v}_{III}$. The resulting shock remains located at the origin at rest as its velocity is ${\bf v} = {\bf v}_{I} + {\bf v}_{II} + {\bf v}_{III} = 0$. 
Hence, we have that the jump of the order parameter at the origin (zero-external fields) for $t = (4/9) \log 2$ can be viewed as a shock wave generated at the earlier ``time" $t =8/27$ at the isolated cusp point $x \simeq -0.008$. This shock will merge at $t=(4/9) \log 2$ with the one propagating from the right along the uniaxiality semi-line.
This analysis also shows that order parameters develop a shock along the uniaxiality semiline $y=x$, $x>0$. Each cusp point corresponds to a gradient catastrophe of order parameters that is formed at infinity ($x\to \infty$) along $y=x$ and propagates towards the origin for $t > 2/9$ reaching the origin at $t = 1/3$. We also observe that, as $t$ increases, the gradient catastrophes generated along the  line $y=x$ develops a jump whose amplitude increases up to a saturation value.  Figure~\ref{Fig:Uniaxial} shows the trajectories of these shocks on the $(x,t)$ plane which separate macroscopic phases associated to isotropic external fields. More precisely, for $x<0$, a transition between the \emph{uniaxial  nematic} $U^+$ and the less ordered \emph{uniaxial para-nematic} $U^-$ phase occurs, while for $x>0$ the system undergoes a uniaxial-to-biaxial phase transition.
In loose terms, the graphs of order parameters intersect along $y=x$ and simultaneously tear up at infinity for $t=2/9$. The rip opens along the line and reaches the origin at $t=1/3$. For instance, Figure~\ref{Fig:uniaxialjump} shows the order parameters $m^{1}$ and $m^{2}$ evaluated along the line $y = 1/3 (-(3/7) + \log(7/4))$ that is such that the cusp point occurs at  the intersection between the first quadrant bisecant and the loop. At $t = 2/7$ both order parameters develop a gradient catastrophe at $x \simeq 0.044$ and for $t>2/7$ the multivalued solution is replaced by the shock solution. Indeeed, Figure~\ref{Fig:cusps} suggests that along a generic trajectory on the $x,y-$plane, order parameters develop a number of gradient catastrophes and shocks at suitable $t$'s given by the number of its intersection with uniaxiality semi-lines and the loop. Figure~\ref{Fig:genericline} shows the order parameter along a the line $y =0.9 x$ intersecting the loop at two points. The two cusp points on the loop are located at $x \simeq -0.01$ and  $x \simeq 0.02$ where both order parameters develop a gradient catastrophe, respectively, at the times $t \simeq 0.294$ and $t \simeq 0.296$. The two shocks generated from these catastrophes merge together at the origin at the time $t = \frac{4}{9} \log 2 \simeq 0.308$ as shown in Figure~\ref{Fig:genericline2} ({\it Left}). Trajectories of these shocks, as shown in  Figure~\ref{Fig:genericline2} ({\it Right}) separate the different macroscopic biaxial phases of the system obtained for non-isotropic external field.

\section{Concluding remarks}
In this paper we put forward a novel approach to the theory of thermotropic phase transitions in nematic liquid crystals within the framework of nonlinear wave equations. We have constructed the exact equations of state for biaxial nematics subject to external fields for the discrete version of the Maier-Saupe model named MS6 model. The statistical mechanical formulation of the theory rests on a general quadratic Hamiltonian modelling interaction between liquid-crystal molecules (endowed with the typical biaxial geometry) and external fields. The explicit expression of the general partition function for a system composed by a finite number of molecules results as the solution to the initial value problem for a diffusive $2+1-$dimensional PDE. In the thermodynamic regime, a suitable asymptotic limit leads to a Hamilton-Jacobi type equation for the free-energy whose solution provides the equations of state for the order parameters. Thermodynamic properties are then determined through the initial conditions which also contain information on the molecular geometry. 

Within the above described conceptual framework, we have obtained the equations of state in explicit analytic form and the occurrence of phase transitions is explained in terms of the mechanism of formation of classical shock waves associated to isotherm and isochamp curves travelling in the space of thermodynamic parameters. 

Detailed study of the equations of state hereby proposed unveils the rich structure of  thermodynamic phase diagrams of the system subject to external fields. It is worth to emphasise that present results are consistent with the numerical analysis obtained for nematics model with external fields considered in~\cite{Frisken} and completes the picture by providing the full description of the critical set in the phase space.

The explicit form obtained for the free-energy in terms of fundamental invariants provides a simple free-energy model for uniaxial nematics. This formula for free-energy captures all relevant features of phase transitions in uniaxial nematics and could be used as thermotropic bulk contribution to general Ginzburgh-Landau models for nematics from a continuum mechanics perspective \cite{BallMajumdar}.

The approach discussed above offers a simple and direct method to study thermodynamics of systems with several scalar order parameters and extends the formulation of statistical mechanics  in terms of nonlinear conservation laws as established in~\cite{BarraMoro, Guerra} and further deployed in \cite{Agliari,Fachechi}. Natural extensions of this theory, currently subject of investigation, are concerned with the study of intrinsically biaxial systems including the role of specific molecules geometry and symmetries and pressure/density effects.

\section*{Acknowledgements}
Authors  would like to thank the London Mathematical Society and GNFM - INdAM for supporting activities that contributed to the research reported in this paper.

\end{document}